\documentclass[
superscriptaddress,
 amsmath,amssymb,
 aps,
]{revtex4-2}

\usepackage{units}
\usepackage{graphicx}
\expandafter\let\csname equation*\endcsname\relax
\expandafter\let\csname endequation*\endcsname\relax

\newcommand{\comment}[1]{}

\begin{document}

\title{Onset of space-charge effects in strong-field photocurrents from nanometric needle tips}

\author{Johannes Sch{\"o}tz}
\affiliation{Department of Physics, Ludwig-Maximilians-Universit\"at Munich, D-85748 Garching, Germany}
\affiliation{Max Planck Institute of Quantum Optics, D-85748 Garching, Germany}
\author{Lennart Seiffert}
\affiliation{Institute for Physics, University of Rostock, D-18051 Rostock, Germany}
\author{Ancyline Maliakkal}
\affiliation{Department of Physics, Ludwig-Maximilians-Universit\"at Munich, D-85748 Garching, Germany}
\affiliation{Max Planck Institute of Quantum Optics, D-85748 Garching, Germany}
\author{Johannes Bl{\"o}chl}
\affiliation{Department of Physics, Ludwig-Maximilians-Universit\"at Munich, D-85748 Garching, Germany}
\affiliation{Max Planck Institute of Quantum Optics, D-85748 Garching, Germany}
\author{Dmitry Zimin}
\affiliation{Max Planck Institute of Quantum Optics, D-85748 Garching, Germany}
\author{Philipp Rosenberger}
\affiliation{Department of Physics, Ludwig-Maximilians-Universit\"at Munich, D-85748 Garching, Germany}
\affiliation{Max Planck Institute of Quantum Optics, D-85748 Garching, Germany}
\author{Boris Bergues}
\affiliation{Department of Physics, Ludwig-Maximilians-Universit\"at Munich, D-85748 Garching, Germany}
\affiliation{Max Planck Institute of Quantum Optics, D-85748 Garching, Germany}
\author{Peter Hommelhoff}
\affiliation{Laser Physics, Department of Physics, Friedrich-Alexander-Universit\"at Erlangen-N\"urnberg, D-91058 Erlangen, Germany}
\author{Ferenc Krausz}
\affiliation{Department of Physics, Ludwig-Maximilians-Universit\"at Munich, D-85748 Garching, Germany}
\affiliation{Max Planck Institute of Quantum Optics, D-85748 Garching, Germany}
\author{Thomas Fennel}
\email{thomas.fennel@uni-rostock.de}
\affiliation{Institute for Physics, University of Rostock, D-18051 Rostock, Germany}
\affiliation{Department of Life, Light and Matter, University of Rostock, D-18059 Rostock, Germany}
\affiliation{Max-Born-Institute, 12489 Berlin, Germany}
\author{Matthias F. Kling}
\email{matthias.kling@lmu.de}
\affiliation{Department of Physics, Ludwig-Maximilians-Universit\"at Munich, D-85748 Garching, Germany}
\affiliation{Max Planck Institute of Quantum Optics, D-85748 Garching, Germany}
	
\begin{abstract}
Strong-field photoemission from nanostructures and the associated temporally modulated currents play a key role in the development of ultrafast vacuum optoelectronics. Optical light fields could push their operation bandwidth into the petahertz domain. A critical aspect for their functionality in the context of applications is the role of charge interactions, including space charge effects. Here, we investigated the photoemission and photocurrents from nanometric tungsten needle tips exposed to carrier-envelope phase-controlled few-cycle laser fields. We report a characteristic step-wise increase in the intensity-rescaled cutoff energies of emitted electrons beyond a certain intensity value. By comparison with simulations, we identify this feature as the onset of charge-interaction dominated photoemission dynamics. Our results are anticipated to be relevant also for the strong-field photoemission from other nanostructures, including photoemission from plasmonic nano-bowtie antennas used in carrier-envelope phase-detection and for PHz-scale devices.
\end{abstract}

\maketitle

\section{Introduction}
The interaction of light with nanostructures exhibits unique features~\cite{stockman2018roadmap}. In plasmonic materials, the excitation of collective oscillations of the electrons with respect to the lattice (i.e. plasmons) leads to a local near-field that can be enhanced up to several orders of magnitude in intensity with respect to the incident field. At the same time, the enhanced fields are confined to the sharp geometrical features of the nanostructure, well below the diffraction limit of light. These nanoplasmonic phenomena have found a vast range of applications (see Ref.~\cite{stockman2018roadmap} for an overview) including biochemical sensing and detection~\cite{Stockman2015NanoplasmonicSensingDetection}, near-field enhanced optical microscopy with nanometer resolution (nanoscopy)~\cite{Gordon2018NearFieldNanoscopy}, surface enhanced Raman spectroscopy~\cite{Langer2019presentSERS} with sensitivity down to the single molecule level~\cite{kneipp1997SERSsingleMolecule}, thermal cancer treatment~\cite{West2007CancerTherapy} and waveguiding of optical energy on the nanometer scale~\cite{Han2013SPPguiding}.

The demonstration that strong-field photoemission from nanometric needle tips~\cite{Hommelhoff2010_TransitionToTunneling, Ropers2010_TransitionToTunneling} and nanospheres can be controlled by the electric field of the driving laser~\cite{ Zherebtsov2011NanosphereCEP, Krueger11attosecondControl}, and that the electron dynamics may be strongly modified by the near-field decay~\cite{Herink2012QuiverQuenching, Echternkamp2016QuiverToSubcycle} has led to the development of attosecond nanophysics as an independent research field~\cite{Hommelhoff2015AttosecondNanophysics, Ciappina2017RepProgPhysAttoNanoscale, Krueger18, dombi2020RevModPstrongnano}. Nanotips illuminated by few-cycle lasers have been used as sources of ultrashort electron pulses for electron microscopy~\cite{muller2014femtosecondElectronSource, vogelsang2018observingElectronSource, feist2018structuralElectronSources} or for the spatially-resolved detection of the carrier-envelope phase (CEP) of a laser beam~\cite{Hommelhoff2017NatPhysCEP}. Significant progress has been made in the direction of lightwave electronics~\cite{goulielmakis2007LightwaveElectronics, Krausz2014AttoMetrology, schotz2019perspective}, i.e. electronics driven by optical fields on the PHz-scale. Based on plasmonic nano-bowtie antennas, the CEP-detection using currents~\cite{Brida2016CEPnanotunneling, Kaertner2017NatPhysCEPcurrentsNanoarray, Kaertner2020onchipCEP}, a potential PHz-scale diode~\cite{Brida2020nanogapcurrent} and recently on-chip field-resolved sampling of optical waveforms~\cite{Bionta2020TipToeChip} have been demonstrated.

It is typically argued that the small dimension of the nano-emission site leads to highly divergent trajectories, justifying the neglect of charge interaction~\cite{Ropers2010_TransitionToTunneling,Lienau2014CEPnanotip,Kaertner2019_vanishingCEP}. On the other hand, for certain parameter ranges the electron kinetic energy spectrum from nanometric needle tips with large apex radii (r$\sim$100\,nm) seems to be completely dominated by Coulomb interaction~\cite{Hiro2016DelayedElectronEmission, yanagisawa2020laser}. Clearly, the miniscule emission area makes nanometric needle tips highly susceptible to space charge effects close to the emission surface. Hence a systematic study of space-charge effects would be highly desirable and of large importance for the ability to maximize the signal in applications related to strong-field emission from nanostructures. Here, we present intensity-dependent photoemission data from nanometric tungsten needle tips in intense few-cycle laser fields recorded via both time-of-flight spectroscopy and photocurrent measurements. By analysis of this data and comparison with simulations, we identify a step-wise increase in the intensity re-scaled cutoff energies of emitted electrons as the onset of charge interaction dominated photoemission dynamics. The results are relevant for many related studies.

\section{Experimental setup}
\begin{figure}[htbp!]
	\centering\includegraphics[width=4in]{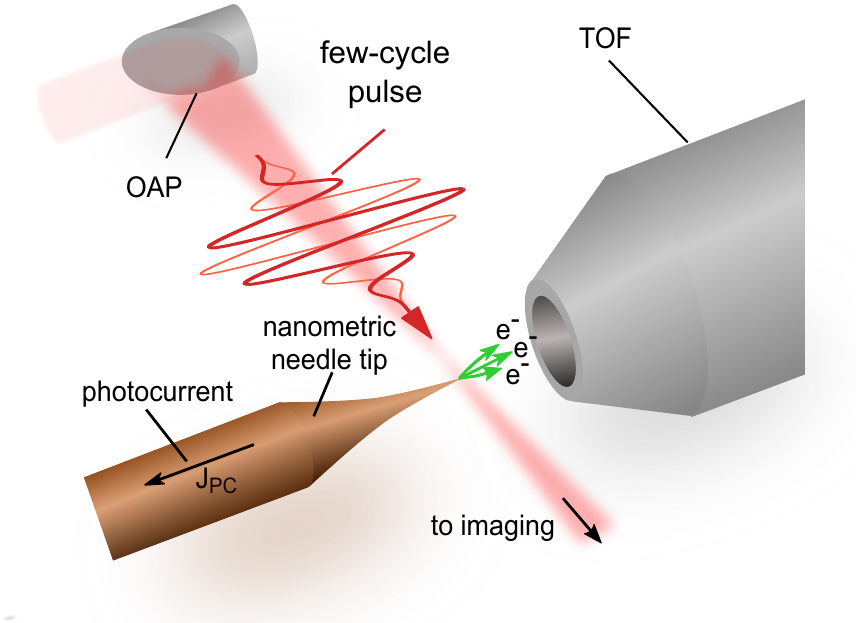}
	\caption[Experimental setup for current detection from a nanometric needle tip]{\label{Nanotip_current_setup} Experimental setup: A linearly polarized few-cycle pulse at 750\,nm with controlled CEP (thick red line showing a cosine pulse for CEP = 0, and thin red line showing a -cosine pulse for CEP = $\pi$) irradiates a tungsten nanometric needle tip. OAP: off-axis parabola, TOF: time-of-flight spectrometer. The CEP and dispersion are adjusted with fused silica wedges (not shown). In addition to the kinetic energy spectrum of emitted electrons measured using the TOF-spectrometer, the photocurrent from the tip is recorded.}
\end{figure}

The experimental setup is illustrated in Fig.\,\ref{Nanotip_current_setup}. We obtained CEP-stable pulses centered at 750\,nm from a hollow-core fiber broadened Ti:sapphire laser system (Spectra Physics, Femtopower Compact Pro HR/CEP) at 10\,kHz repetition rate. The pulses were compressed with a set of chirped mirrors (Ultrafast Innovations, PC70) to 4.5\,fs in full-width-at-half-maximum (FHWM) of the intensity envelope, which was determined via the Dispersion Scan (d-scan) technique~\cite{Miranda12}. A pair of fused silica wedges was used to control the dispersion and the CEP of the pulses. The few-cycle pulses with controlled CEP were focused onto a nanometric tungsten needle tip using an off-axis parabola (OAP, f=10\,cm). The needle tip was produced using wet chemical etching~\cite{Ibe1990TipEtching}, resulting in a typical apex radius of (40$\pm$20)\,nm. The emitted photoelectrons were detected using a time-of-flight (TOF)-spectrometer (Stefan Kaesdorf, ETF10) and a time-to-digital-converter (FAST ComTec, P7889). Due to the vacuum-requirements for the electron spectra measurements and the microchannel-plate detector of the spectrometer, the setup was placed in a vacuum chamber. The photoemitted electrons also resulted in a photocurrent from the nanotip. The photocurrent was amplified using a low-noise high-gain transimpedance amplifier (FEMTO, DLPCA-200) and detected using a lock-in amplifier (Z\"urich Instruments, HF2LI). For the measurements of the CEP-dependence, the CEP of consecutive laser pulses was flipped between $\phi_\mathrm{0}$ and $\phi_\mathrm{0}+\pi$ using the acousto-optic programmable dispersive filter (Fastlite Dazzler) of the Ti:sapphire laser system.

The digital lock-in amplifier was capable of demodulating the signal input at several frequencies in parallel. For the first frequency, we chose the laser repetition rate $f_\mathrm{rep}=10\,\mathrm{kHz}$ which resulted in a signal proportional to the total photocurrent. The second frequency was the CEP-flipping frequency $f_\mathrm{CEP}$ which was set to 5\,kHz. We observed the best signal to noise performance at the maximum gain of $10^9\,\mathrm{V/A}$ of the transimpedance amplifier even though the nominal gain bandwidth ($f_\mathrm{-3dB}$) was below 5\,kHz, which slightly reduced the gain at 10\,kHz. The conversion factor between signal amplitude and electrons per shot is approximately given by $0.6\,\mathrm{electrons}\cdot \mathrm{shot}^{-1}\mu\mathrm{V}^{-1}$. Our experiments are conducted in the multi-electron emission regime with up to several hundred electrons per shot. Similar conditions were found for the nano-bowtie current experiments driven at MHz repetition rates~\cite{Kaertner2020onchipCEP}. Their study indirectly indicates $10^3$ to $10^4$ electrons per shot per nanostructure, based on the CEP-current per shot and nanostructure (0.11\,electrons) and the ratio of CEP-current to total current ($10^{-4}$ to $10^{-5}$). The results of our studies are thus of direct importance also in related experiments where currents from nanostructures are measured.

\section{Results and discussion}
\begin{figure}[htbp!]
	\centering\includegraphics[width=0.5\columnwidth]{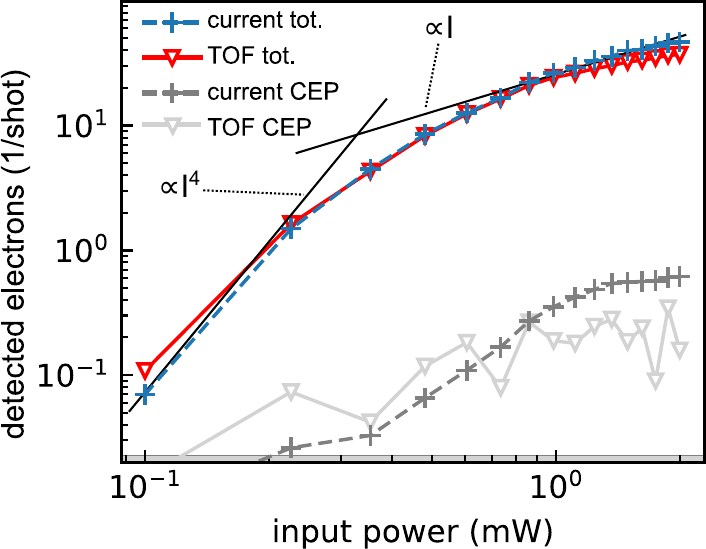}
	\caption[Energy-resolved electron spectra and charge interaction effects]{\label{Nanotip_current_TOFintensity} Evidence of charge interaction in the recorded electron yield and photocurrent measurement: TOF (red) and current (blue) count rate together with the detected CEP dependence (light and dark gray, respectively). The current curves have been scaled by a factor 1/70. The change in slope towards linear emission indicates substantial charge interaction. The black lines serve as a guide to the eye.}
\end{figure}

The connection between the photocurrent emitted from and flowing through the tip (and measured with the lock-in amplifier as discussed above - short 'the photocurrent') and the TOF measurements is established in Fig.\,\ref{Nanotip_current_TOFintensity}, which shows the intensity dependence of the number of detected electrons. In our experiments, we observed no significant damage to the nanotip during the measurement even at the highest intensities, as confirmed by the repeatability of the measurements. Both photocurrent (blue crosses and dashed line) and TOF (red triangles and solid line) measurements show a very similar evolution. However, in the photocurrent, approximately a factor 70 more electrons are detected. The main reason is likely the partial transmission of the TOF spectrometer due to the acceptance angle of only around 2.5$^\circ$. We intentionally did not make use of the lens in the TOF spectrometer in order to avoid distortions of the spectra. From a rough estimate considering the emission angle~\cite{Ropers2010_TransitionToTunneling} we would expect a factor of 40, which is close to our observation. This illustrates the advantage of the photocurrent approach to capture all emitted electrons if this is required to characterize the dynamics which may allow shorter measurement times. The total number of detected electrons per shot in the photocurrent increases from below 5 at the lowest input power of 0.1\,mW to above 3000 at 2\,mW. The photocurrent may include slow, potentially thermally emitted electrons that are not resolved in the TOF measurement. Most importantly, the change of the slope in Fig.\,\ref{Nanotip_current_TOFintensity} towards a linear emission regime is interpreted as a signature for the onset of substantial charge interaction.

Figure \ref{Nanotip_current_TOFintensity} also shows the amplitude of the CEP-dependent modulation of the photocurrent (dark gray crosses and dashed line) and the electron yield from the TOF measurement (light gray triangles and solid line) by scanning the CEP over 20 values between 0 and 2$\pi$. Note that the statistical error in the TOF measurements is significantly higher (not shown for the sake of clarity of the figure): on the order of the oscillations between consecutive datapoints, since the measurements are affected by slow laser power drifts and fluctuations unlike the photocurrent that is measured with a lock-in amplifier. A ratio of CEP-dependent modulation signal to total count rate of around $10^{-2}$ is measured, which is between one and two orders of magnitude above what has been reported in other studies using nano-bowtie structures and MHz repetition rate sources~\cite{Kaertner2017NatPhysCEPcurrentsNanoarray}. The reason could be twofold. Firstly, it has been reported that in MHz repetition rate nanotip experiments the strong-field emission is suppressed due to accumulative heating~\cite{bionta2016RepRatedependentEmission}. Secondly, we use shorter input pulses and non-resonant field enhancement that translates into enhanced near-fields with a pulse duration similar to the incident pulse.

\begin{figure}[htbp!]
	\centering
	\includegraphics[width=1.0\columnwidth]{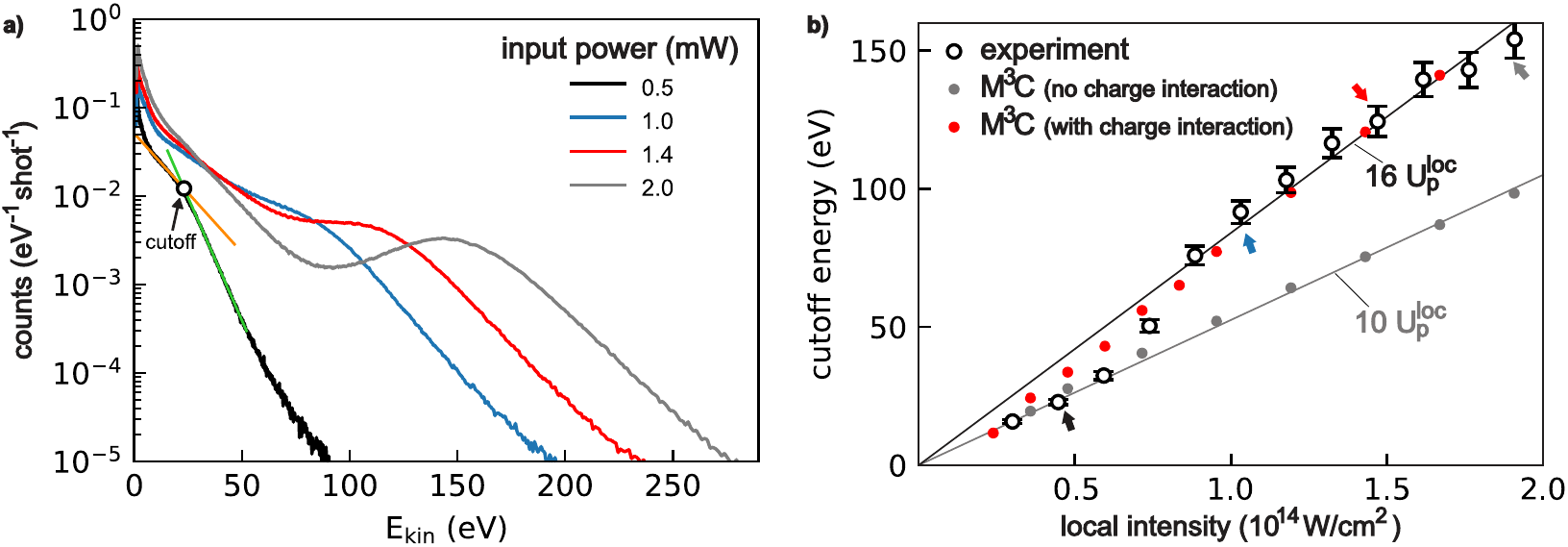}
	\caption[Energy-resolved electron spectra and charge interaction effects]{\label{Nanotip_current_TOFintensity_scaling} Impact of charge interaction on the cutoff energies of recollision electrons emitted from a nanometric tungsten needle tip: a)~CEP-averaged electron energy spectra for different laser input powers (as indicated). The cutoff energies are defined via the intersection of linear fits of the data below (orange line) and above (green line) the apparent cutoff of the respective  recollision plateaus (here visualized only for the lowest power). b)~Cutoff energies as function of near-field intensity (black circles). Colored arrows indicate the cutoffs corresponding to the spectra in a). Gray and red symbols show the cutoffs extracted from semi-classical M$^3$C simulations performed for the experimental parameters when ex- and including charge interaction, respectively. Gray and black lines visualize the 10\,$U_\text{p}^\text{loc}$ cutoff law as expected from the simple mans model and a rescaled 16\,$U_\text{p}^\text{loc}$ cutoff law. The scaling factor between input laser power and local intensity is deduced from matching the lowest three experimental cutoff values to the 10\,$U_\text{p}^\text{loc}$-cutoff law.}
\end{figure}

CEP-averaged photoelectron energy spectra measured by the TOF for various local intensities are shown in Fig.\,\ref{Nanotip_current_TOFintensity_scaling}\,a. At low intensities (black curve), a low-energy peak connected to a plateau is observed, reminiscent of the direct electron and rescattering contributions of strong-field photoemission corresponding to the 2\,$U_\text{p}^\text{loc}$ and 10\,$U_\text{p}^\text{loc}$-cutoffs, similar to e.g. Ref.~\cite{Krueger11attosecondControl}. Note that here $U_\text{p}^\text{loc}$ is the local ponderomotive potential of the enhanced near-field. Beyond the plateau, the spectra decrease rapidly but still extend to quite high kinetic energy. This is measurable thanks to the high dynamic range of the spectrometer. For each data set, the value of the cutoff of the plateau is evaluated by fitting straight lines to the logarithmic plot of the data below (orange) and above (green) the apparent cutoff and determining their intersection point, as illustrated here for the lowest input power shown. As the laser power is increased, the cutoff of the plateau evolves into a peak structure (cf. gray curve in Fig.\,\ref{Nanotip_current_TOFintensity_scaling}\,a). The formation of the peak is consistent with the typical structure of elastic backscattering at higher intensity, where the underlying classical trajectory picture becomes increasingly appropriate~\cite{Becker_JPB51_2018}.

Clear indications for the onset of considerable charge interaction effects are found in both the power scaling of the yield in Fig.~\ref{Nanotip_current_TOFintensity}, which no longer reflects the typical strongly non-linear yield increase with intensity but grows nearly linearly, and the intensity scaling of the plateau cutoff (black circles) shown in Fig.\,\ref{Nanotip_current_TOFintensity_scaling}b. For the lowest intensities, a linear scaling of the cutoff corresponding to the 10\,$U_\text{p}^\text{loc}$-cutoff law (solid gray line) is found. At a local intensity of about $\unit[7.5\times10^{13}]{W/cm^2}$ a step-wise increase of the cutoffs is observed, followed by an again nearly linear scaling corresponding to a modified cutoff-law of about 16\,$U_\text{p}^\text{loc}$ (solid black line). Both effects, i.e. the nearly linear yield evolution with laser power and the increased cutoff are consistent with previous results found for dielectric nanospheres~\cite{Sussmann2015DirectionalEmission, seiffert_JMO64_2017}. According to these studies, ionization is limited by charge separation at the surface. The capacitor-like field formed by the released electrons and the residual ions remaining in the nanostructure quenches tunnel ionization, leading to a nearly linear growth of the electron yield with power. The additional energy gain is attributed to modified recollison dynamics due to the trapping field resulting from the charge separation and space charge repulsion among the electrons in the departing bunches~\cite{Sussmann2015DirectionalEmission, seiffert_JMO64_2017}.

To substantiate that the observed cutoff enhancement is caused by charge interaction, we performed semi-classical trajectory simulations for the experimental parameters. We adapted our Mean-Field Mie Monte-Carlo (M$^3$C) model, previously utilized to study strong-field photoemission and attosecond streaking at dielectric nanospheres~\cite{Sussmann2015DirectionalEmission,rupp_JMO64_2017, Max2020SphereMetallization, seiffert2017attosecond}, for the description of electron emission from a metallic nanotip. The details of the original nanosphere model are described in~\cite{seiffert2018PhD}. In brief, the near-field of a dielectric sphere is evaluated as a combination of the linear polarization field caused by an incident laser pulse and additional non-linear contributions from charge interaction that we treat as a mean-field in electrostatic approximation. The latter includes both Coulomb interactions among and the additional sphere polarization caused by the free charges, i.e. liberated electrons and residual ions. Electron trajectories are generated by Monte-Carlo sampling of ADK-type tunneling rates~\cite{Ammosov_JETP64_1986} and are propagated in the local near-field by integrating classical equations of motion. For electrons propagating in the material, we account for elastic electron-atom collisions and inelastic collisions (impact ionization) via respective scattering cross-sections. To describe the photoemission from a metallic nanotip, we considered three key modifications. First, we allow tunneling only from one half-sphere to mimic the apex of the nanotip. Second, we account for the emerging image charges within the metal in the describtion of the mean-field. Third, we use a Fowler-Nordheim type tunneling rate and consider a workfunction of \unit[6.5]{eV} for the oxidized tungsten nanotip similar to earlier studies~\cite{Krueger11attosecondControl, Wachter12, wachter2014PhD}. We note that the rate was scaled with a linear factor for best agreement with the experimental data.

The cutoff energies extracted from M$^3$C simulations with the mean-field turned off and on are shown in Fig.\,\ref{Nanotip_current_TOFintensity_scaling}\,b) as gray and red symbols, respectively. The simulations excluding the mean-field predict cutoffs following the 10\,$U_\text{p}^\text{loc}$-law. When including the mean-field, cutoffs around 10\,$U_\text{p}^\text{loc}$ are only obtained at low intensities where ionization and thus charge-interaction-induced modifications of the local near-fields are weak. At higher intensities, the cutoff converges to 16\,$U_\text{p}^\text{loc}$ in close agreement with the experiment. Although the transition between the two cutoff laws is less rapid than in the measured data, the semiclassical model captures the main trends and thus confirms that the observed step-like feature is a clear signature for the onset of charge interaction dominated photoemission.

So far, most systematic studies focusing on charge interaction in strong-field photoemission were performed on isolated nanospheres~\cite{Zherebtsov2011NanosphereCEP,Suessmann2013PhD,Sussmann2015DirectionalEmission,Max2020SphereMetallization}. In these works, several effects could be identified by thorough analysis of the experimental results and extensive numerical simulations. Despite the differences in the strong-field photoemission process between nanometric metallic needle tips and dielectric nanospheres, the similarity of the observed charge-interaction effects, the strong reduction of the nonlinearity of the photoemission process and the increase of the cutoff energy, are striking.

Our measurements indicate that charge interaction starts to affect the electron dynamics above a near-field intensity of $\unit[7.5\times10^{13}]{W/cm^2}$ or around 1000 e$^{-}$/shot for 4.5~fs pulses at 750\,nm and tungsten needle tips with a tip radius of around 40\,nm. While the strong-field tunneling photocurrent experiments on nano-bowties and triangles~\cite{Brida2016CEPnanotunneling, Kaertner2019_vanishingCEP, Brida2020PRB_attodiode, Brida2020nanogapcurrent, Bionta2020TipToeChip} focus mainly on the CEP-dependent current which is on the order of one electron per shot, the total number of charges per shot is typically several orders of magnitude higher~\cite{Kaertner2017NatPhysCEPcurrentsNanoarray, Kaertner2019_vanishingCEP, Kaertner2020onchipCEP} and therefore in a similar regime as for our experiment. Our study also shows that despite the clear charge interaction, waveform-dependent photocurrents can be measured, which is important for the development of applications of field-controlled currents. The discussion of charge interaction given here is therefore of high relevance to these studies as well. Additionally, the higher kinetic energies of electrons due to the charge interaction could prove useful in future applications.

\section{Conclusions}
We have investigated the photoemission and photocurrents from nanometric tungsten needle tips in CEP-controlled few-cycle laser fields. For multi-electron emission, we identified two regimes, where initially the cutoff energies of emitted electrons closely follow what is expected from near-field enhanced backscattering dynamics. At the onset of charge interactions becoming dominant in the emission process, we observed a step-wise increase in the cutoff energies.
The results are relevant also for the strong-field electron emission from other nanostructures, including studies where ultrafast currents from plasmonic bow-tie nanostructures have been used for CEP-detection and PHz-scale optoelectronic devices.

\section*{Acknowledgements}
We acknowledge fruitful discussions with Matthew Weidman and support in the fabrication of the nanometric needle tips by Mohammed Qahosh, Markus Bohn, and Hirofumi Yanagisawa. T.F. acknowledges computing time provided by the North German Supercomputing Alliance (HLRN, ID mvp00017).

\section*{Funding}
  http://dx.doi.org/10.13039/501100001659,"Deutsche Forschungsgemeinschaft", via SPP QUTIF, LMUexcellent, and projects 389759512 and 398382624; http://dx.doi.org/10.13039/501100000781,"European Research Council", via FETopen PetaCOM and FETlaunchpad FIELDTECH; http://dx.doi.org/10.13039/501100004189,"Max-Planck-Gesellschaft", via the IMPRS for Advanced Photon Science and the Max-Planck Fellow program.


\end{document}